\documentclass[prb]{revtex4}
\usepackage{graphicx}
\begin{document}
\title{Reversal modes in arrays of interacting magnetic 
Ni nanowires: Monte Carlo simulations and scaling technique
}
\author{M. Bahiana}
\affiliation{Instituto de F\'{\i}sica, Universidade Federal do Rio de Janeiro\\
Caixa Postal 68528, Rio de Janeiro, RJ, Brazil, 21945-970}
\author{S. Allende}
\affiliation{Departamento de F\'{\i}sica, Universidad de Santiago\\
Casilla 307, Santiago 2, Chile}
\author{F. S. Amaral}
\affiliation{Instituto de F\'{\i}sica, Universidade Federal do Rio de Janeiro\\
Caixa Postal 68528, Rio de Janeiro, RJ, Brazil, 21945-970}
\author{D. Altbir}
\affiliation{Departamento de F\'{\i}sica, Universidad de Santiago\\
Casilla 307, Santiago 2, Chile}
\date{\today}
\begin{abstract}
The effect of dipolar interactions in hexagonal arrays of Ni nanowires has been investigated 
by means of Monte Carlo simulations combined with a scaling technique, which allows the investigation of the internal structure of the wires. A strong dependence of the coercivity and remanence on the distance between wires has been observed. At intermediate packing densities the coercivity exhibits a maximum, higher than the non-interacting value. This behavior, experimentally observed, has been explained on grounds of the interwire dipolar interactions.  Also, different reversal modes of the magnetization have been identified. 
\end{abstract}
\pacs{75.75.+a,75.60.Ej,75.60.Jk}

\maketitle
\section{Introduction}
In recent years, great deal of attention has been focused on the
study of regular arrays of magnetic particles with dimensions in
the nanometer range. These particles have potential applications
in nonvolatile magnetic memory devices or high-resolution magnetic
field sensors\cite{chapman} and arrays of discrete patterned magnetic
elements, such as magnetic wires, rings and dots, have been
proposed as a new generation of ultrahigh density patterned
magnetic storage media\cite{ross1}. Experimental and theoretical
results over the past years show that there are many factors, 
such as geometry, anisotropy and magnetic interactions among the particles, 
that influence the magnetic behavior of the systems. The
question of how the interelement spacing affects the magnetic
properties of an array is relevant in understanding and
interpreting experimental results since this spacing can influence
both the magnetization reversal mechanism and the internal
magnetic domain structures. The effect of interparticle
interactions is in general complicated by the fact that the
dipolar fields depend upon the magnetization state of each
element, which in turn depend upon the fields due to adjacent
elements. Therefore, modeling of such systems is often subject to
strong simplifications like, for example, considering monodomain particles. In the specific case of wires, Sampaio {\it et
al.}\cite{sampaio} have described an array of microwires as a one-dimensional
array of Ising-like magnetic  moments subject to an anisotropy
field, representing the wire shape anisotropy, and with the 
dipolar interaction taken into account as a field depending on the orientation of the
participating magnetic moments. Hysteresis curves
with some of the features observed in experiments were obtained by
Monte Carlo simulations. A more realistic model was
presented in \cite{sampaio2} to describe one and two-dimensional
arrays of microwires. In this case the magnetic moments were allowed
to point at any direction on the plane, and dipolar interactions
have been directly calculated. 
This approach provides a
description limited to very long wires in a weakly interacting
regime, excluding the exploration of many interesting issues.
Models for wires with non-uniform magnetization have been
restricted to micromagnetic calculations
\cite{hertel,hertel2,liviu}. The introduction of internal degrees
of freedom enables mapping the field in the vicinity of the wire
\cite{liviu} and identifying a corkscrew reversal mode
\cite{hertel}. 

In this article we develop Monte Carlo simulations
for an array of nanowires in which the internal structure of each
wire is taken into account. We focus on arrays of  nanowires
created by electro-deposition of nickel in porous alumina
membranes \cite{nielschAPL}. This fabrication technique produces
hexagonal arrays of nanowires with long-range ordering, and well
controlled center to center distance ($D$), diameter ($d$) and
length ($\ell$). Typical values are $d=$10 to 100 nm, $\ell=0.1$
to 1$\mu$m and $D=30$ to 100 nm. More frequently studied are wires
with aspect-ratios $\ell/d>10$ in order to enforce the bit
character of individual elements, so we also consider nanowires
with this characteristic. Experimentally it is possible to observe
that the coercive field strongly depends on the ratio $d/D$,
indicating that the reversal process is greatly affected by
magnetostatic interaction among wires \cite{vazquezpirota}. This
effect is observed in several experiments, in which the hysteresis
curves for  membranes with different packing densities are
measured \cite{nielschAPL,vazquezpirota,sellmyer2001}.
Wires in the above mentioned range of sizes  have at least $10^8$
atoms, and since dipolar interactions must be considered in those
systems, numerical simulations at the atomic level are out of
reach with the present computational facilities. In order to
circumvent this difficulty, we make use of a scaling technique
presented  by d'Albuquerque {\it et al.}\cite{scaling1}, which was
applied to the calculation of the phase diagram of cylindrical
particles. In this approach the number of particles is reduced to
a value suitable for numerical calculations, which decreases the
dipolar field felt by each particle. The exchange coupling constant is then
scaled down in order to keep the correct
balance between magnetostatic and exchange energies, responsible
for domain formation and reversal mechanisms. 
This technique, combined with standard Monte Carlo simulations 
has been used in the study of nanometric elements, providing results otherwise unattainable 
with this approach \cite{fastmontecarlo,mejia2005}. 

Modeling an
array with macroscopic dimensions, even with the scaling
procedure, is not possible due to the large number of wires. 
In order to extract information about the reversal process taking into account the interaction with neighboring wires and, at the same time, use reasonable computational time, we have studied an hexagonal cell with seven wires, considering the central wire, which interacts with the full number of first neighboring wires, as representative of a typical element of a macroscopic array. 
Monte Carlo simulations were used to calculate
the hysteresis curve for cells with different interwire spacing. The behavior
of the coercivity and the reversal modes for weak and strong interacting limits were
analyzed.

\section{Model}
The internal energy, $E_{tot}$, of a  wire array with $N$ magnetic
moments can be written as
\begin{equation}
E_{tot} =
\sum_{i=1}^N\sum_{j > i}^N\left( E_{ij}-J \widehat{\mu }_{i}\cdot
\widehat{\mu }_{j}\right)
+ E_K + E_H ,
\label {Etot}
\end{equation}
where $E_{ij}$ is the dipolar energy given by
\begin{equation}
E_{ij}=\frac{\overrightarrow{\mu }_{i}\cdot \overrightarrow{\mu }_{j}-3(%
\overrightarrow{\mu }_{i}\cdot \widehat{n}_{ij})(\overrightarrow{\mu }%
_{j}\cdot \widehat{n}_{ij})}{r_{ij}^{3}} \; .
\label{eq:Edip}
\end{equation}
$r_{ij}$ is the distance between the magnetic moments
$\overrightarrow{\mu }_{i}$ and $\overrightarrow{\mu }_{j},$ and
$\widehat{n}_{ij}$ the unit vector along the direction that
connects the two magnetic moments. $J$ is the exchange coupling
constant between nearest neighbors, and
 $\widehat{\mu }_{i}$ is the unit vector
along the direction of $\overrightarrow{\mu }_{i}$.
 $E_K$ is a cubic
crystalline anisotropy term which can be written as
$E_{K}=K\sum_{i=1}^{N}\left[\alpha _{i}^{2}\beta _{i}^{2}+\beta _{i}^{2}\gamma
_{i}^{2}+\gamma _{i}^{2}\alpha _{i}^{2}\right] $, where $(\alpha
_{i},\beta _{i},\gamma _{i})$ are the direction cosines of
$\overrightarrow{\mu }_{i}$ referred to the cube axis\cite{Kittel},
 and
$E_H=-\sum_{i=1}^N\overrightarrow{\mu}_i\cdot\overrightarrow{H}$
is the contribution of the external field.

In order to compare our simulations with experimental results on granular Ni systems,
we have considered $|\overrightarrow{\mu}_i|=\mu =0.615\mu_{B}$,
the lattice parameter $a_{0}=3.52$ $\AA$, $K=2\times 10^5$erg/cm$^3$ and $J=1600$ kOe/$\mu_B$\cite{Kittel}.
The wires have diameter $d = 30$ nm, length $\ell=1$ $\mu$m, and were
 built along the [110] direction of a fcc lattice comprising
  about $6\times 10^9$ atoms. In order to reduce the
number of interacting atoms, we make use of the scaling technique
presented  before \cite{scaling1}, applied to the calculation of
the phase diagram of cylindrical particles of height $\ell$ and
diameter $d$. The authors showed that such diagram is equivalent
to the one for a smaller particle with $d^\prime = d \chi^\eta$
and $\ell^\prime = \ell \chi^\eta$, being $\chi<1$ and $\eta
\approx 0.56$, if the exchange constant is also scaled as
$J^\prime = \chi J$. It has also been showed \cite{fastmontecarlo} that
the scaling relations can be used together with Monte Carlo
simulations to obtain a general magnetic state of a nanoparticle.
We use this idea starting from the desired value for the total
number of interacting particles we can deal with. Based on the
computational facilities currently available, we have estimated
that a total $N\approx 3500$ to be a reasonable value. With
this in mind we have obtained the value $\chi=8\times 10^{-4}$,
that leads to wires with 504 atoms each.

In what follows we simulate hysteresis curves at temperature $T=300$ K, using
the scaling technique described above. It is important to observe
that when measuring a hysteresis loop, the value of the
coercivity is affected by the rate at which the external field is
varied. Similarly, in  simulations of that curve the number of
Monte Carlo steps for each value of the field is a critical issue
to be defined. We have followed the procedure used by many
authors, in which  the number of Monte Carlo steps for a particular
case is varied until fair agreement with experimental results is
obtained \cite{rest3,alloy,neqmc}. Then, the number of Monte Carlo steps
is kept fixed and all other variables can be changed. Monte Carlo
simulations were carried out using Metropolis algorithm with local
dynamics and single-spin flip methods\cite{binder}. The new
orientation of the magnetic moment was chosen arbitrarily with a
probability $p=\mbox{min}[1,\exp (-\Delta E/k_B T)]$, where
$\Delta E$ is the change in energy due to the reorientation of the
magnetic moment, and $k_B$ is the Boltzmann constant. 
One interesting point to be considered is the effect of scaling on
temperature at which the simulations are carried out. Since our goal is
to obtain hysteresis curves, we need to figure out how the scaling is
affecting the transition between metastable states. The energy landscape of
the system is rather complicated due to the dipolar interaction, but in the vicinity
of each local minimum we can analyze the transitions as regulated by
energy barriers of the form $K_eV$ where $K_e$ is an effective
anisotropy constant which takes into account several energy
contributions, and $V$ is the volume of the particle. 
Thermal activated transitions
naturally lead to the definition of a blocking
temperature $T_B \propto K_eV$ \cite{cullity}, so we
use this to relate temperature and size.
 In order to keep thermal activation
 process invariant under the scaling transformation, the energy barriers must also be invariant, therefore, temperature should scale as the volume, that is, $T^\prime=\chi^{3\eta}T$. From now on, all results refer to the scaled system.

The hysteresis loops were simulated with the external field in the direction of the wire axis.  The initial state had the field $H = 2.0$ kOe, higher
than the saturation field, and a configuration in which all
the magnetic moments were aligned with the external field. The field
was then linearly decreased at a rate of 300  Monte Carlo steps for  $\Delta H=0.01$ kOe.
In this way, to go from saturation to the coercive field about 120.000 MC steps are needed. 
The values of coercivity correspond to an average over, at least, 10 independent realizations.

\section{Results and Discussion}
\begin{figure}
\begin{center}
\includegraphics{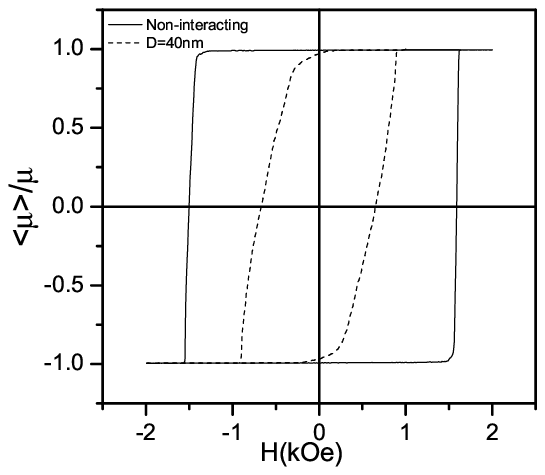}
\end{center}
\caption{Hysteresis curves for an isolated wire,
 and for the central wire of an hexagonal array
 of seven, with interwire distance $D=40$ nm.
\label{fig:hist}}
\end{figure}

Our main concern in this work is to investigate the role of
dipolar interactions in  wire arrays, specially its effect in the
coercivity.  Figure ~\ref{fig:hist} shows the hysteresis
loops for an isolated nanowire and for the central wire of a cell
 with  spacing $D=40$ nm. Comparing
the curves we can immediately conclude that interaction affects
the reversal processes not only in respect to the coercivity
values, but also to the shape of the curve. The  loop for the
isolated wire has a 100\% squareness, while the one for the
interacting central wire, only 45\%. Although not shown in the
figure, the curve for the central wire of the $D=100$ nm array
almost coincides with the one for an isolated wire. 
Similar results have been found by other authors.
For example, regular arrays of monodisperse columns have been modeled by Samwell {\it et al.} \cite{samwelJ3M1992} and 
Yshii and Sato \cite{ishiiJ3M1989}. Both papers report analytical calculations of an internal field parameter for nanowires in a membrane. Considering magnetostatic interactions as the only cause of shear in the hysteresis loop, the authors find good agreement with experimental values. Results reported by Sorop {\it et al.} \cite{soropPRB2003} for Fe nanowires embedded in nanoporous alumina templates reinforce this idea. By examining the wire morphology, and varying the temperature, the authors have discarded the influence of these factors in the shape of the hysteresis loop. The squareness of the hysteresis loop has also been examined by Hwang {\it et al.} \cite{hwangIEEE2000}, who fitted experimental curves for arrays of cylindrical Ni particles with a deterministic model in which the cylinders are represented by a single magnetic moment. The arrays have one adjustable parameter, the standard deviation $\sigma$ for the switching field distribution. By comparing simulations of hysteresis curves for systems of interacting particles and $\sigma=0$, and non-interacting particles and $\sigma\neq 0$, with experimental curves, they were able to conclude that the shear observed is due to interaction among particles. 
%

\begin{figure}[h]
\begin{center}
\includegraphics{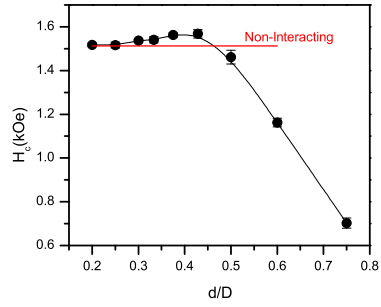}
\end{center}
\caption{Coercivity of the central wires of an hexagonal array of seven, as a function of the interwire distance,
 $D$. The wires have diameter $d= 30$ nm, and length $\ell = 10^3$ nm. The solid line is a guide to the eye.\label{fig:hc}}
\end{figure}

\begin{figure}
\begin{center}
\includegraphics{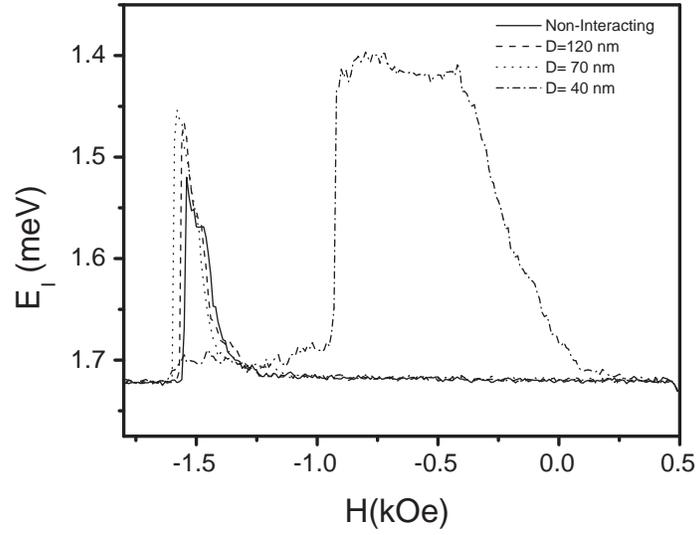}
\end{center}
\caption{Internal dipolar energy  for the central wire in arrays
with $D= 40$, 70 and 120 nm, and for an isolated wire, along the
hysteresis cycle. \label{fig:eforma}}
\end{figure}

The effect of interwire spacing on the
 coercivity can be examined in Fig. \ref{fig:hc}, where
 its value is plotted as a function of $d/D$, for values
  of $D$ corresponding to almost non-interacting wires, $D=150$ nm,
 up to strongly interacting ones, for $D=40$ nm. 
The same behavior was observed in a square arrangement, but with a less pronounced maximum.
For comparison, the value
  obtained for an isolated wire is represented by the
 horizontal line.
The large-$D$ regime coincides with the non-interacting limit, but
it is interesting to note that the transition to the
strongly interacting regime involves a maximum in coercivity. Since
we are looking at the central wire, the curve in Fig.~\ref{fig:hc}
reflects the reversal order of the group of wires.  In the limit
of non-interacting wires, $D\geq 150$ nm ($d/D \leq 0.20$) all of them
reverse basically at the same time, for 100 nm $\geq D \geq 70$ nm ($0.30\leq d/D \leq 0.43$) the
central wire is the last one to revert, and for $60$ nm $\geq D \geq 40$ nm
($0.50\leq d/D \leq 0.75$) it is the first one in the reversal process.
This increase in stability has also been reported by Hertel \cite{hertel}. The author
has performed micromagnetic simulations of hexagonal arrays of nanowires with fixed
geometric parameters. Using our notation, his system is composed by wires with $\ell=1\mu$m, 
$d=40$nm and $D=100$nm, leading to $d/D=0.4$, well within the maximum coercivity region.
Hertel examines the effect of increasing the number of wires in the array, as a form of increasing 
the interaction among wires, and observes that the the reversal of some them occurs because the
stray field of neighboring wires adds to the external field and
leads to a higher field to which the magnetic moments are
effectively exposed as compared to a single nanowire.
On the other hand, those wires remaining with magnetization
antiparallel to the field are confronted with the stray
field of the reversed wires which is oriented opposite to the
external field thus reducing the local field. His conclusion is that, in this case,
saturation is reached at higher field strength compared to a
single wire.
The reversal process is regulated, in first order, by the
internal dipolar energy of the wire, $E_I$. This energy corresponds to the dipolar interaction between the magnetic moments within each wire. 
Fig.~\ref{fig:eforma} shows the behavior of $E_I$ along the
hysteresis curve, for
 the central wire of arrays with
 $D=40$, 70 and 120 nm, and for an isolated wire.
 For $D>70$ nm the reversal is fast,
resulting in a sharp peak in the energy curve, localized at the
coercive field. For $D=40$ nm, the reversal starts at zero field,
and has a duration about fives times larger. 

\begin{figure}
\begin{center}
\includegraphics{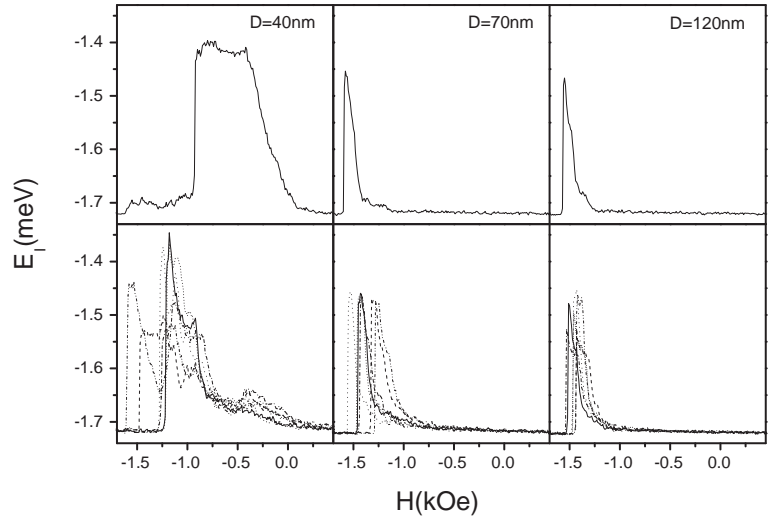}
\end{center}
\caption{Internal dipolar  energy $E_A$ for individual wires along the hysteresis cycle. The upper line corresponds to the central wire, while the lower
one shows the reversal curves for each of the six external wires.\label{fig:arrayeforma}}
\end{figure}

The complexity of the reversal process in strongly interacting
arrays is evident when one compares the value of $E_I$ for each wire, 
for different values of $D$,
along the hysteresis cycle. The upper curves in Fig.~\ref{fig:arrayeforma}  illustrate
$E_I$ of the central wire for $D= 40$, 70 and 120
nm, while the lower curves depict $E_I$ of the six external wires of the array, for the
same values of $D$. For $D=120$ nm, the process consists of a sequence of
sharp reversals within an interval of 0.3 kOe. For $D=70$ nm we
observe that a variation of about 0.5 kOe is needed to reverse all
wires, but it is still possible to identify the reversal of each
individual wire. The situation for the strongly interacting array,
with $D = 40$ nm, is quite different. Each reversal curve
has a complicated structure and the superposition is large. Also,
the peaks are wider and span a large interval of field values. 
The reversal of the whole
array involves a variation of about 1.6 kOe, and only the
reversion of the central wire can be well separated from the
others, acting as a trigger to the reversion of the surrounding
wires.

\begin{figure}
\begin{center}
\includegraphics{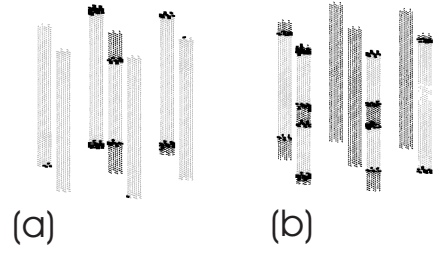}
\end{center}
\caption{Reversal process in an array with $D= 40$ nm along the hysteresis curve, for  two
different moments. Magnetic moments aligned opposite to the field are represented in light grey, while those already reversed appear in dark grey. Black regions represent the domain walls. (a) Reversion
starts at the central wire, where domain walls nucleate at the
extremes and propagate
 towards the center. (b) In two of the border wires
there is also the nucleation of domain walls in the central part.
These wall propagate towards the tips, merging with the ones generated there. \label{fig:rede}}
\end{figure}

Examining the internal structure of each wire, we notice another effect from the interaction.
Reversion in isolated wires occurs via nucleation of domain walls at the wire tips, that
propagate and merge near the center as found also in micromagnetic simulations \cite{hertel}.
In the $D=40$ nm array we have also observed the nucleation of domain walls 
at the center  of the wire,
propagating towards the tips and merging with the walls coming from there. Fig.~\ref{fig:rede}
shows two moments of the reversal process for such array. In Fig.~\ref{fig:rede}(a) the
central wire has already started to revert and has two domain walls
traveling towards the center. A snapshot taken later  (Fig.~\ref{fig:rede}(b))
shows three wires completely reverted, and two of the outer wires with
nucleation of domain walls in the central part also.

In order to better understand the appearance of the maximum at $D
= 70$ nm, the relative stability of possible wire configurations must be investigated.
For this purpose we have calculated  magnetization and energy of the
hexagonal array in the absence of external field, assuming that the wires were
saturated with magnetization along the wire axis. 
For $D = 40$ nm, the dipolar energy of the central wire is
$-1.55$ meV, while the energy of each outer wire is $-1.60$
meV. Since the central wire has a higher energy, its reversion is more likely to occur.
For $D = 70$ nm the energies become $-1.66$ meV for the
central wire and $-1.67$ meV for the outer ones. 
In this case the energy difference is not large enough to make
the central wire considerably less stable. Actually, as some magnetic
moments in the outer wires acquire components transverse to the wire axis,
the central wire has its energy lowered, becoming more stable.
 The reversal process starts easily and simultaneously in two opposite
external wires, separated by a distance equal to $2D$. This intermediate
configuration is very favorable since comprises six antiferromagnetic bonds between
nearest neighboring wires, and four between next-nearest neighboring wires.
The reversal process continues with two other pairs of opposite external wires, 
being the central wire the last one to reverse,  exhibiting a coercivity that is
even larger that the one for a non interacting system. For $D
> 150$ nm the array may be considered as a non-interacting one,
with all the wires reversing
essentially at the same time since they are identical and independent.

\section{Conclusions}
In this paper we have used an scaling technique combined with
Monte Carlo simulations in order to investigate the reversal of an
hexagonal array of seven wires. The possibility of such
 a scale reduction increases considerably the applicability of numerical simulations
 to material science in general. This method allows us  to consider
the internal structure of each wire during the reversal process.
With the proposed simulation scheme we were able to reproduce
experimental results for Ni nanowires, that is the decrease in
remanence and coercivity as interaction becomes stronger \cite{vazquez1}.
The shear observed in the hysteresis curve can be attributed to interaction among wires,
 a result supported by independent simulations and analytical calculations by other authors.
The existence of a maximum in our coercivity curve (Fig.~\ref{fig:hc}) resides in
the definition of a particular reversal order of the wires determined mainly by 
the dipolar interaction. We believe that the maximum observed in experimental curves
\cite{sellmyer2000} is originated by a similar process. The positional disorder of the wires, which is always present in real macroscopic arrays, may create local cells generating blocking of innermost wires
 Since our wires had no internal
disorder, we can discard the influence of such effect in the
behavior of coercivity. 
Our results show that strongly interacting systems experience reversal process much slower than non interacting wires.

\section{Acknowledgements}
This work was partially supported by  FONDECYT under grants
1050013, and 7050273, and from  the Millennium Science Nucleus
"Condensed  Matter Physics" P02-054F, and CNPq, PIBIC/UFRJ, FAPERJ, PROSUL
Program, and Instituto de Nanoci\^{e}ncias/MCT of Brazil. One of
the authors, S. A., ackowledge the support from CONICYT Ph.D.
Program Fellowships, and Direcci\'on de Postgrado, Universidad de
Santiago de Chile and PROSUL, which finance his stay at
Universidade Federal do Rio de Janeiro.

%
%

\newpage

\end{document}